\documentclass[showpacs,a4paper,twocolumn,nofootinbib,superscriptaddress]{revtex4-2}
\usepackage{graphicx}
\usepackage{newpxmath,newpxtext}
\usepackage{graphicx}
\usepackage{epsfig}
\usepackage[normalem]{ulem}
\usepackage{xcolor}

\newcommand{\der}[2]{\dfrac{d#1}{d#2}}
\newcommand{\pder}[2]{\dfrac{\partial#1}{\partial#2}}
\newcommand{\pdder}[3]{\dfrac{\partial^2 #1}{\partial #2 \partial #3}}

\newcommand{\dder}[2]{\dfrac{\delta#1}{\delta#2}}
\newcommand{\pdot}[1]{\dot{\partial}_{#1}}

\newcommand{\D}{\mathcal{D}}
\newcommand{\Gd}{\mathcal{G}}

\newcommand{\R}{\mathcal{R}}
\newcommand{\de}{\mathrm{d}}

\newcommand{\lin}{\\[7pt]}
\usepackage{tensor}

\begin{document}
	
	\title{Applications of the Schwarzschild-Finsler-Randers model}
	
	% Author 1
	\author{E. Kapsabelis}
	\email{manoliskapsabelis@yahoo.gr}
	\affiliation{Section of Astrophysics, Astronomy and Mechanics, Department of 
		Physics, National and Kapodistrian University of Athens, Panepistimiopolis 15784, Athens, Greece}
	
	% Author 2
	\author{A. Triantafyllopoulos}
	\email{alktrian@phys.uoa.gr}
	\affiliation{Section of Astrophysics, Astronomy and Mechanics, Department of 
		Physics, National and Kapodistrian University of Athens, Panepistimiopolis 15784, Athens, Greece}
	
	% Author 3
	\author{S. Basilakos}
	\email{svasil@academyofathens.gr}
	\affiliation{Academy of Athens, Research Center for Astronomy and Applied Mathematics, Soranou Efessiou 4, 115 27 Athens, Greece}
	\affiliation{National Observatory of Athens, Lofos Nymfon, 11852, Athens, Greece}
	
	% Author 4
	\author{P. C. Stavrinos}
	\email{pstavrin@math.uoa.gr}
	\affiliation{Department of Mathematics, National and Kapodistrian University of 
		Athens,	Panepistimiopolis 15784, Athens, Greece}
	
	\begin{abstract}
		In this article, we study further applications of the Schwarzschild-Finsler-Randers (SFR) model which was introduced in a previous work \cite{Triantafyllopoulos:2020vkx}. In this model, we investigate curvatures and the generalized Kretschmann invariant which plays a crucial role for singularities. In addition, the derived path equations are used for the gravitational redshift of the SFR-model and these are compared with the GR model. Finally, we get some results for different values of parameters of the generalized photonsphere of the SFR-model and we find small deviations from the classical results of general relativity (GR) which may be ought to the possible Lorentz violation effects. 
	\end{abstract}
	
	
	\maketitle
	
	\section{Introduction}
	The last decade has seen a rapid increase of Finsler and Finsler-like geometries and their applications to gravitation and cosmology with appreciable results in the scientific community. We quote some relevant works which have contributed in the development of applications of Finsler and Finsler-like geometries to the gravitational field theory and cosmology \cite{Miron:1994nvt,stavrinos2005,Vacaru:2005ht,Gibbons:2007iu,Stavrinos:2006rf,Chang:2007vq,Kostelecky:2008be,Kouretsis:2008ha,Kouretsis:2010vs,Mavromatos:2010jt,Kostelecky:2011qz,Skakala:2010hw,Kouretsis:2012ys,mavromatos2012,Stavrinos:2012kv,Vacaru:2010fi,Vacaru:2010rd,Pfeifer:2011xi,AlanKostelecky:2012yjr,Stavrinos:2012ty,Basilakos:2013hua,Basilakos:2013ij,Silva:2013xba,Stavrinos:2013neo,Li:2014taa,Javaloyes:2013ika,Foster:2015yta,Minguzzi:2014aua,Fuster:2015tua,Stavrinos:2016xyg,Hohmann:2016pyt,Ootsuka:2016zru,Antonelli:2018fbv,Javaloyes:2018lex,Triantafyllopoulos:2018bli,Pfeifer:2019wus,Minas:2019urp,Ikeda:2019ckp,Chowdhury:2019gbq,Triantafyllopoulos:2020vkx,Hohmann:2020mgs,Lobo:2020qoa,Torri:2020fao,Triantafyllopoulos:2020ogl,Relancio:2020mpa,Li:2020fvh,Caponio:2020ofw,Elbistan:2020mca,Stavrinos:2021ygh}.
	
	Finsler geometry is a dynamical metric geometry depending on position and direction or dynamical coordinates on a tangent or fiber bundle of a differentiable manifold. This type of geometry can also be connected to Lorentz violation investigations of the standard model extension (SME) \cite{Girelli:2006fw,Torri:2019gud} and in the context of local anisotropy \cite{Gibbons:2007iu,Kouretsis:2008ha,Kostelecky:2011qz,AlanKostelecky:2012yjr}. Moreover, Finsler-like geometries breaking the local four dimensional Lorentz invariance can be considered as a possible alternative direction for investigating physical models with both local anisotropy and violation of local spacetime symmetries \cite{Vacaru:2005ht}.
	
	A significant class of Finslerian spacetime is the Finsler-Randers (FR) spacetime proposed by Randers \cite{randers1940}.
	An FR space has a metric function of the form
	\begin{equation}\label{lagrangian}
		F(x,y) = (-a_{\mu\nu}(x)y^{\mu}y^{\nu})^{1/2} + u_{\alpha}y^{\alpha}    
	\end{equation}
	where $u_{\alpha}$ is a covector with $||u_{\alpha}||\ll 1$, $y^{\alpha}=\frac{dx^{\alpha}}{d\tau}$ and $a_{\mu\nu}(x)$ is a Riemannian metric for which the Lorentzian signature $(-,+,+,+)$ has been assumed and the indices $\mu, \nu, \alpha$ take the values $0,1,2,3$. The geodesics of this space can be produced by \eqref{lagrangian} and the Euler-Lagrange equations. If $u_{\alpha}$ denotes a force field $f_{\alpha}$ and $y^{\alpha}$ is substituted with $d x^{\alpha}$ then $f_{\alpha}dx^{\alpha}$ represents the spacetime effective energy produced by the anisotropic force field $f_{\alpha}$, therefore equation \eqref{lagrangian} is written as
	\begin{equation}\label{lagrangian2}
		F(x,dx) = \left(-a_{\mu\nu}(x)dx^{\mu}dx^{\nu}\right)^{1/2} + f_{\alpha}dx^{\alpha}    
	\end{equation}
	The integral $\int_{a}^{b}F(x,dx)$ represents the total work that some particle needs to move along a path.
	
	The length of a curve $c$ in the FR space is given by 
	\begin{equation}
		l(c) = \int^{1}_{0}F(x,\dot{x})d\tau    
	\end{equation}
	where $\dot{x} = \frac{dx}{d\tau}$ and $\tau$ is affine parameter.
	
	An FR cosmological model was introduced and studied in \cite{stavrinos2005}, \cite{Stavrinos:2006rf}. In this case,
	by considering the metric of the FRW cosmological model instead of $a_{\mu\nu}(x)$ in \eqref{lagrangian2} we get
	\begin{equation}
		a_{\mu\nu}(x) = \mathrm{diag}\left[-1,\frac{a^2}{1-\kappa r^{2}},a^{2}r^{2}, a^{2}r^{2}\sin^{2}\theta\right]   
	\end{equation}
	and we obtain a Finsler-Randers cosmology. From \eqref{lagrangian2} we can notice that an FR spacetime shows a motion of the FRW model with a produced work which comes from the second term (one-form). This form of metric provides a dynamic effective structure in spacetime. More investigations about this model can be found in the following articles \cite{Chang:2007vq,Stavrinos:2012kv,Stavrinos:2016xyg,Stavrinos:2002rc,Papagiannopoulos:2017whb,Silva:2015ptj,Chaubey:2018wph,Raushan:2020mkh,Papagiannopoulos:2020mmm,Silva:2020tqr,Hama:2021frk}.
	
	By using a Schwarzschild metric in \eqref{lagrangian}, we obtain a Schwarzschild - Randers spacetime \cite{Triantafyllopoulos:2020vkx}.
	\begin{align}\label{lagrangian3}
		F(x,y) = &\left[-\left(1-\frac{R_s}{r}\right)(y^t)^2 + \frac{(y^r)^2}{1-\frac{R_s}{r}} + r^2 (y^\theta)^2 \right. +\nonumber\\ &  + r^2 \sin^{2}\theta\, (y^\phi)^2 \Bigg]^{1/2} + u_{\alpha}y^{\alpha}    
	\end{align}
	From \eqref{lagrangian3}, we can also see that the Schwarzschild-Randers metric has a dynamical second term.
	
	Finsler and Finsler-Randers spacetimes can give an effective description of fermion particles with CPT-odd Lorentz violating terms in the SME framework \cite{AlanKostelecky:2012yjr,Silva:2015ptj,Colladay:2019lig}.
	
	In this work, we elaborate some fundamental results of the SFR model and compare them with the corresponding ones of GR. We prove that the gravitational redshift predicted from our model remains invariant compared with the one of GR. Nevertheless, in the case of photon sphere, we find infinitesimal deviations from GR which may be ought to the small anisotropic perturbations coming from Lorentz violation effects. In addition, in our generalized metric space, we calculate the Kretschmann invariants of the model and we find that the generalized second Kretschmann invariant $K_V$ provides more information for singularities with additional degrees of freedom.
	
	This article is organized as follows: In sec. \ref{sec: Basic structure} we give some basic elements from the geometry of SFR. In sec. \ref{sec: Curvatures} we present the curvatures and the field equations. In sec. \ref{sec: Paths}, \ref{sec: Energy}, \ref{sec: Redshift} and \ref{sec: Photonsphere} we give some applications of the SFR model including paths, energy, gravitational redshift and photonsphere. Finally, in the last section \ref{sec: Conclusion} we summarize the results of our work.
	
	\section{Basic structure of the model}\label{sec: Basic structure}
	
	In this section, we briefly present the underlying geometry of the SFR gravitational model, as well as the field equations for the SFR metric. The solution of these equations for this metric is presented at the end of the section. An extended study of this model can be found in \cite{Triantafyllopoulos:2020vkx,Triantafyllopoulos:2020ogl}.
	
	The Lorentz tangent bundle $TM$ is a fibered 8-dimensional manifold with local coordinates $\{x^\mu,y^\alpha\}$ where the indices of the $x$ variables are $\kappa,\lambda,\mu,\nu,\ldots = 0,\ldots,3$ and the indices of the $y$ variables are  $\alpha,\beta,\ldots,\theta = 4,\ldots,7$.
	The tangent space at a point of $TM$ is spanned by the so called adapted basis
		$\{E_A\} = \,\{\delta_\mu,\dot\partial_\alpha\} $ with
		\begin{equation}
			\delta_\mu = \dfrac{\delta}{\delta x^\mu}= \pder{}{x^\mu} - N^\alpha_\mu(x,y)\pder{}{y^\alpha} \label{delta x}
		\end{equation}
		and
		\begin{equation}
			\dot \partial_\alpha = \pder{}{y^\alpha}
		\end{equation}
		where $N^\alpha_\mu$ are the components of a nonlinear connection $N=N^{\alpha}_{\mu}(x,y)\,\de x^{\mu}\otimes \pdot{\alpha}$. %\frac{\partial}{\partial y^{\alpha}}$.
		
		The nonlinear connection induces a split of the total space $TTM$ into a horizontal distribution $T_HTM$ and a vertical distribution $T_VTM$. The above-mentioned split is expressed with the Whitney sum:
		\begin{equation}
			TTM = T_HTM \oplus T_VTM
		\end{equation}
		The anholonomy coefficients of the nonlinear connection are defined as
		\begin{equation}\label{Omega}
			\Omega^\alpha_{\nu\kappa} = \dder{N^\alpha_\nu}{x^\kappa} - \dder{N^\alpha_\kappa}{x^\nu}
	\end{equation}
	A Sasaki-type metric $\Gd$ on $TM$ is:
	\begin{equation}
		\mathcal{G} = g_{\mu\nu}(x,y)\,\mathrm{d}x^\mu \otimes \mathrm{d}x^\nu + v_{\alpha\beta}(x,y)\,\delta y^\alpha \otimes \delta y^\beta \label{bundle metric}
	\end{equation}
	We define the metrics $g_{\mu\nu}$ and $v_{\alpha\beta}$ to be pseudo-Finslerian.
	
	A pseudo-Finslerian metric $ f_{\alpha\beta}(x,y) $ is defined as one that has a Lorentzian signature of $(-,+,+,+)$ and that also obeys the following form:
	\begin{align}
		f_{\alpha\beta}(x,y) = \pm\frac{1}{2}\pdder{F^2}{y^\alpha}{y^\beta} \label{Fg}
	\end{align}
	where the function $F$ satisfies the following conditions \cite{Skakala:2010hw}:
	\begin{enumerate}
		\item $F$ is continuous on $TM$ and smooth on  $ \widetilde{TM}\equiv TM\setminus \{0\} $ i.e. the tangent bundle minus the null set $ \{(x,y)\in TM | F(x,y)=0\}$ . \label{finsler field of definition}
		\item $ F $ is positively homogeneous of first degree on its second argument:
		\begin{equation}
			F(x^\mu,ky^\alpha) = kF(x^\mu,y^\alpha), \qquad k>0 \label{finsler homogeneity}
		\end{equation}
		\item The form 
		\begin{equation}
			f_{\alpha\beta}(x,y) = \pm\dfrac{1}{2}\pdder{F^2}{y^\alpha}{y^\beta} \label{finsler metric} 
		\end{equation}
		defines a non-degenerate matrix: \label{finsler nondegeneracy}
		\begin{equation}
			\det\left[f_{\alpha\beta}\right] \neq 0 \label{finsler nondegenerate}
		\end{equation}
	\end{enumerate}
	where the plus-minus sign in  \eqref{Fg} is chosen so that the metric has the correct signature.
	
	In this work, we will follow the model presented in \cite{Triantafyllopoulos:2020vkx}. The metric $g_{\mu\nu}$ is the classic Schwarzschild one:
	\begin{align}\label{Schwarzchild}
		&g_{\mu\nu}\de x^\mu  \de x^\nu \nonumber\\
		&\quad = -fdt^2 + \frac{dr^2}{f} + r^2 d\theta^2 + r^2 \sin^{2}\theta\, d\phi^2
	\end{align}
	with $f=1-\frac{R_s}{r}$ and $R_s=2GM$ the Schwarzschild radius (we assume units where $c=1$).
	
	Hereafter, we consider an $\alpha$-Randers type metric as the one in rel.\eqref{lagrangian} which is distinguished from the $\beta$-Randers type metric that is investigated in the SME \cite{Kostelecky:2011qz,AlanKostelecky:2012yjr,Silva:2013xba,Foster:2015yta}.
	
	The metric $v_{\alpha\beta}$ is derived from a metric function $F_v$ of the $\alpha$-Randers type:
	\begin{equation}\label{RandersL}
		F_v = \sqrt{-g_{\alpha\beta}(x)y^\alpha y^\beta} + A_\gamma(x) y^\gamma
	\end{equation}
	where $g_{\alpha\beta}=g_{\mu\nu}\tilde\delta^{\mu}_{\alpha}\tilde\delta^{\nu}_{\beta}$ is the Schwarzschild metric and $A_{\gamma}(x)$ is a covector which expresses a deviation from general relativity, with $|A_\gamma(x)|\ll 1$. The nonlinear connection will take the form:
	\begin{equation}\label{Nconnection}
		N^\alpha_\mu = \frac{1}{2}y^\beta g^{\alpha\gamma}\partial_\mu g_{\beta\gamma}
	\end{equation}
	The metric tensor $v_{\alpha\beta}$ of \eqref{RandersL} is derived from \eqref{Fg} after omitting higher order terms $O(A^2)$:
	\begin{equation}\label{vab}
		v_{\alpha\beta}(x,y) = g_{\alpha\beta}(x) + w_{\alpha\beta}(x,y)   
	\end{equation}
	where 
	\begin{align}\label{wab}
		w_{\alpha\beta} = & \, \frac{1}{\tilde{a}}(A_{\beta}g_{\alpha\gamma}y^\gamma + A_{\gamma}g_{\alpha\beta}y^\gamma + A_{\alpha}g_{\beta\gamma}y^\gamma) \nonumber\\
		& + \frac{1}{\tilde{a}^3}A_{\gamma}g_{\alpha\epsilon}g_{\beta\delta}y^\gamma y^\delta y^\epsilon
	\end{align}
	with $\tilde{a} = \sqrt{-g_{\alpha\beta}y^{\alpha}y^{\beta}}$.
	The total metric defined in the previous steps is called the Schwarzschild-Finsler-Randers (SFR) metric.
	
	In this work, we consider a distinguished connection ($d-$connection) $ {D} $ on $TM$. This is a linear connection with coefficients $\{\Gamma^A_{BC}\} = \{L^\mu_{\nu\kappa}, L^\alpha_{\beta\kappa}, C^\mu_{\nu\gamma}, C^\alpha_{\beta\gamma} \} $ which preserves by parallelism the horizontal and vertical distributions:
		\begin{align}
			{D_{\delta_\kappa}\delta_\nu = L^\mu_{\nu\kappa}(x,y)\delta_\mu} \quad &,\quad D_{\pdot{\gamma}}\delta_\nu = C^\mu_{\nu\gamma}(x,y)\delta_\mu \label{D delta nu} \lin
			{D_{\delta_\kappa}\pdot{\beta} = L^\alpha_{\beta\kappa}(x,y)\pdot{\alpha}} \quad &, \quad D_{\pdot{\gamma}}\pdot{\beta} = C^\alpha_{\beta\gamma}(x,y)\pdot{\alpha} \label{D partial b}
		\end{align}
		From these, the definitions for partial covariant differentiation follow as usual, e.g. for $X \in TTM$ we have the definitions for covariant h-derivative
		\begin{equation}
			X^A_{|\nu} \equiv D_\nu\,X^A \equiv \delta_\nu X^A + L^A_{B\nu}X^B \label{vector h-covariant}
		\end{equation}
		and covariant v-derivative
		\begin{equation}
			X^A|_\beta \equiv D_\beta\,X^A \equiv \dot{\partial}_\beta X^A + C^A_{B\beta}X^B \label{vector v-covariant}
		\end{equation}
		The $d-$connection is metric-compatible when the following conditions are met:
		\begin{equation}
			D_\kappa\, g_{\mu\nu} = 0, \quad D_\kappa\, v_{\alpha\beta} = 0, \quad D_\gamma\, g_{\mu\nu} = 0, \quad D_\gamma\, v_{\alpha\beta} = 0
		\end{equation}
		A $d-$connection can be uniquely defined given that the following conditions are satisfied:
			\begin{itemize}
				\item The $d-$connection is metric compatible
				\item Coefficients $L^\mu_{\nu\kappa}, L^\alpha_{\beta\kappa}, C^\mu_{\nu\gamma}, C^\alpha_{\beta\gamma} $ depend solely on the quantities $g_{\mu\nu}$, $v_{\alpha\beta}$ and $N^\alpha_\mu$
				\item Coefficients $L^\mu_{\kappa\nu}$ and $ C^\alpha_{\beta\gamma} $ are symmetric on the lower indices, i.e.  $L^\mu_{[\kappa\nu]} = C^\alpha_{[\beta\gamma]} = 0$
		\end{itemize}
		We use the symbol $\mathcal D$ instead of $D$ for a connection satisfying the above conditions, and call it a canonical and distinguished $d-$connection.
	The coefficients of canonical and distinguished $d-$connection are
	%can be found in \cite{miron-watanabe-ikeda 1987}:
	\begin{align}
		L^\mu_{\nu\kappa} & = \frac{1}{2}g^{\mu\rho}\left(\delta_kg_{\rho\nu} + \delta_\nu g_{\rho\kappa} - \delta_\rho g_{\nu\kappa}\right) \label{metric d-connection 1}  \\
		L^\alpha_{\beta\kappa} & = \dot{\partial}_\beta N^\alpha_\kappa + \frac{1}{2}v^{\alpha\gamma}\left(\delta_\kappa v_{\beta\gamma} - v_{\delta\gamma}\,\dot{\partial}_\beta N^\delta_\kappa - v_{\beta\delta}\,\dot{\partial}_\gamma N^\delta_\kappa\right) \label{metric d-connection 2}  \\
		C^\mu_{\nu\gamma} & = \frac{1}{2}g^{\mu\rho}\dot{\partial}_\gamma g_{\rho\nu} \label{metric d-connection 3} \\
		C^\alpha_{\beta\gamma} & = \frac{1}{2}v^{\alpha\delta}\left(\dot{\partial}_\gamma v_{\delta\beta} + \dot{\partial}_\beta v_{\delta\gamma} - \dot{\partial}_\delta v_{\beta\gamma}\right) \label{metric d-connection 4}
	\end{align}
	
	Curvatures and torsions on $TM$ can be defined by the multilinear maps:
		\begin{equation}
			\mathcal{R}(X,Y)Z = [\mathcal{D}_X,\mathcal{D}_Y]Z - \mathcal{D}_{[X,Y]}Z \label{Riemann tensor TM}
		\end{equation}
		and
		\begin{equation}
			\mathcal{T}(X,Y) = \mathcal{D}_XY - \mathcal{D}_YX - [X,Y] \label{torsion TM}
		\end{equation}
		where $X,Y,Z \in TTM$.
		We use the following definitions for the curvature components \cite{Miron:1994nvt,Vacaru:2005ht}:
		\begin{align}
			\mathcal{R}(\delta_\lambda,\delta_\kappa)\delta_\nu = R^\mu_{\nu\kappa\lambda}\delta_\mu \label{R curvature components} \lin
			\mathcal{R}(\delta_\lambda,\delta_\kappa)\pdot{\beta} = R^\alpha_{\beta\kappa\lambda}\pdot{\alpha}\lin
			\mathcal{R}(\pdot{\gamma},\delta_\kappa)\delta_\nu = P^\mu_{\nu\kappa\gamma}\delta_\mu \lin
			\mathcal{R}(\pdot{\gamma},\delta_\kappa)\pdot{\beta} = P^\alpha_{\beta\kappa\gamma}\pdot{\alpha}\lin
			\mathcal{R}(\pdot{\delta},\pdot{\gamma})\delta_\nu = S^\mu_{\nu\gamma\delta}\delta_\mu\lin
			\mathcal{R}(\pdot{\delta},\pdot{\gamma})\pdot{\beta} = S^\alpha_{\beta\gamma\delta}\dot{\partial}_\alpha \label{S curvature components}
		\end{align}
		In addition, we use the following definitions for the torsion components:
		\begin{align}
			\mathcal{T}(\delta_\kappa,\delta_\nu) = & \mathcal{T}^\mu_{\nu\kappa}\delta_\mu + \mathcal{T}^\alpha_{\nu\kappa}\pdot{\alpha} \label{torsion components 1} \lin
			\mathcal{T}(\pdot{\beta},\delta_\nu) = & \mathcal{T}^\mu_{\nu\beta}\delta_\mu + \mathcal{T}^\alpha_{\nu\beta}\pdot{\alpha} \label{torsion components 2} \lin
			\mathcal{T}(\pdot{\gamma},\pdot{\beta}) = & \mathcal{T}^\mu_{\beta\gamma}\delta_\mu + \mathcal{T}^\alpha_{\beta\gamma}\pdot{\alpha} \label{torsion components 3}
	\end{align}
	The h-curvature tensor of the $d-$connection in the adapted basis and the corresponding h-Ricci tensor have, respectively, the components given from \eqref{R curvature components}:
	\begin{align}
		& R^\mu_{\nu\kappa\lambda} = \delta_\lambda L^\mu_{\nu\kappa} - \delta_\kappa L^\mu_{\nu\lambda} + L^\rho_{\nu\kappa}L^\mu_{\rho\lambda} - L^\rho_{\nu\lambda}L^\mu_{\rho\kappa} + C^\mu_{\nu\alpha}\Omega^\alpha_{\kappa\lambda} \label{R coefficients 1}\lin
		& R_{\mu\nu} = R^\kappa_{\mu\nu\kappa} =  \delta_\kappa L^\kappa_{\mu\nu} - \delta_\nu L^\kappa_{\mu\kappa} + L^\rho_{\mu\nu}L^\kappa_{\rho\kappa} - L^\rho_{\mu\kappa}L^\kappa_{\rho\nu}\nonumber\\
		& \quad\quad\quad\quad\quad\quad + C^\kappa_{\mu\alpha}\Omega^\alpha_{\nu\kappa} \label{d-ricci 1}
	\end{align}
	The v-curvature tensor of the $d-$connection in the adapted basis and the corresponding v-Ricci tensor have, respectively, the components \eqref{S curvature components}:
	\begin{align}
		S^\alpha_{\beta\gamma\delta} & = \pdot{\delta} C^\alpha_{\beta\gamma} - \pdot{\gamma}C^\alpha_{\beta\delta} + C^\epsilon_{\beta\gamma}C^\alpha_{\epsilon\delta} - C^\epsilon_{\beta\delta}C^\alpha_{\epsilon\gamma} \label{S coefficients 2} \lin
		S_{\alpha\beta} & = S^\gamma_{\alpha\beta\gamma} = \pdot{\gamma}C^\gamma_{\alpha\beta} - \pdot{\beta}C^\gamma_{\alpha\gamma} + C^\epsilon_{\alpha\beta}C^\gamma_{\epsilon\gamma} - C^\epsilon_{\alpha\gamma}C^\gamma_{\epsilon\beta} \label{d-ricci 4}
	\end{align}
	The generalized Ricci scalar curvature in the adapted basis is defined as
	\begin{equation}
		\R = g^{\mu\nu}R_{\mu\nu} + v^{\alpha\beta}S_{\alpha\beta} = R+S \label{bundle ricci curvature}
	\end{equation}
	where
	\begin{align}
		R=g^{\mu\nu}R_{\mu\nu} \quad,\quad
		S=v^{\alpha\beta}S_{\alpha\beta} \label{hv ricci scalar}
	\end{align}

	A Hilbert-like action on $TM$ can be defined as
		\begin{equation}\label{Hilbert like action}
			K = \int_{\mathcal N} d^8\mathcal U \sqrt{|\Gd|}\, \R + 2\kappa \int_{\mathcal N} d^8\mathcal U \sqrt{|\Gd|}\,\mathcal L_M
		\end{equation}
		for some closed subspace $\mathcal N\subset TM$, where $|\Gd|$ is the absolute value of the metric determinant, $\mathcal L_M$ is the Lagrangian of the matter fields, $\kappa$ is a constant and
		\begin{equation}
			d^8\mathcal U = \de x^0 \wedge \ldots \wedge\de x^3 \wedge \de y^4 \wedge \ldots \wedge \de y^7
		\end{equation}
		Variation with respect to $g_{\mu\nu}$, $v_{\alpha\beta}$ and $N^\alpha_\kappa$ leads to the following field equations \cite{Triantafyllopoulos:2020ogl} (see Appendix \ref{sec: Variation} for more details):
	\begin{align}
		& \overline R_{\mu\nu} - \frac{1}{2}({R}+{S})\,{g_{\mu\nu}} \nonumber\\
		& \quad + \left(\delta^{(\lambda}_\nu\delta^{\kappa)}_\mu - g^{\kappa\lambda}g_{\mu\nu} \right)\left(\mathcal D_\kappa\mathcal T^\beta_{\lambda\beta} - \mathcal T^\gamma_{\kappa\gamma}\mathcal T^\beta_{\lambda\beta}\right)  = \kappa {T_{\mu\nu}} \label{feq1}\\
		& S_{\alpha\beta} - \frac{1}{2}({R}+{S})\,{v_{\alpha\beta}} \nonumber\\
		& \quad + \left(v^{\gamma\delta}v_{\alpha\beta} - \delta^{(\gamma}_\alpha\delta^{\delta)}_\beta \right)\left(\mathcal D_\gamma C^\mu_{\mu\delta} - C^\nu_{\nu\gamma}C^\mu_{\mu\delta} \right) = \kappa {Y_{\alpha\beta}} \label{feq2}\\
		& g^{\mu[\kappa}\pdot{\alpha}L^{\nu]}_{\mu\nu} +  2 \mathcal T^\beta_{\mu\beta}g^{\mu[\kappa}C^{\lambda]}_{\lambda\alpha} = \frac{\kappa}{2}\mathcal Z^\kappa_\alpha \label{feq3}
	\end{align}
	with
	\begin{align}
		T_{\mu\nu} &\equiv - \frac{2}{\sqrt{|\Gd|}}\frac{\Delta\left(\sqrt{|\Gd|}\,\mathcal{L}_M\right)}{\Delta g^{\mu\nu}} = - \frac{2}{\sqrt{-g}}\frac{\Delta\left(\sqrt{-g}\,\mathcal{L}_M\right)}{\Delta g^{\mu\nu}}\label{em1}\\
		Y_{\alpha\beta} &\equiv -\frac{2}{\sqrt{|\Gd|}}\frac{\Delta\left(\sqrt{|\Gd|}\,\mathcal{L}_M\right)}{\Delta v^{\alpha\beta}}  = -\frac{2}{\sqrt{-v}}\frac{\Delta\left(\sqrt{-v}\,\mathcal{L}_M\right)}{\Delta v^{\alpha\beta}}\label{em2}\\
		\mathcal Z^\kappa_\alpha &\equiv -\frac{2}{\sqrt{|\Gd|}}\frac{\Delta\left(\sqrt{|\Gd|}\,\mathcal{L}_M\right)}{\Delta N^\alpha_\kappa} = -2\frac{\Delta\mathcal{L}_M}{\Delta N^\alpha_\kappa}\label{em3}
	\end{align}
	where $\mathcal L_M$ is the Lagrangian of the matter fields, $\delta^\mu_\nu$ and $ \delta^\alpha_\beta$ are the Kronecker symbols, $|\Gd|$ is the absolute value of the determinant of the total metric \eqref{bundle metric}, and
	\begin{equation}\label{torsion}
		\mathcal{T}_{\nu\beta}^{\alpha} = \pdot{\beta} N_{\nu}^{\alpha} - L_{\beta\nu}^{\alpha}
	\end{equation}
	are torsion components, where $L_{\beta\nu}^{\alpha}$ is defined in \eqref{metric d-connection 2}. From the form of \eqref{bundle metric} it follows that $\sqrt{|\Gd|} = \sqrt{-g}\sqrt{-v}$, with $g, v$ the determinants of the metrics $g_{\mu\nu}, v_{\alpha\beta}$ respectively.
	
	Solving the above equations to first order in $A_\gamma(x)$ in vacuum ($T_{\mu\nu} = Y_{\alpha\beta} = \mathcal Z^\kappa_\alpha = 0$), we get  \cite{Triantafyllopoulos:2020vkx}:
	\begin{equation}\label{Asolution}
		A_\gamma(x) = \left[\tilde A_0 \left|1-\frac{R_S}{r} \right|^{1/2}, 0, 0, 0 \right]
	\end{equation}
	with $\tilde A_0$ a constant.
	\section{Curvatures and generalized Kretschmann invariants}\label{sec: Curvatures}
	It is useful to calculate invariants of the metrics $g_{\mu\nu}$ and $v_{\alpha\beta}$ so that we can get a better understanding for the behaviour of the solution in specific points of $TM$. Specifically, any point in $TM$ where these invariants diverge can be considered singular, namely, a point where our geometrical model breaks down.
	
	We consider invariants constructed from contractions of the curvature tensor on $TM$ and the metric. The nonvanishing curvature components of the distinguished canonical connection are given by relations \eqref{R curvature components}-\eqref{S curvature components}:
	\begin{align}
		R^\mu_{\nu\kappa\lambda} & =  \delta_\lambda L^\mu_{\nu\kappa} - \delta_\kappa L^\mu_{\nu\lambda} + L^\rho_{\nu\kappa}L^\mu_{\rho\lambda} - L^\rho_{\nu\lambda}L^\mu_{\rho\kappa} + C^\mu_{\nu\alpha}\Omega^\alpha_{\kappa\lambda} \\
		R^\alpha_{\beta\kappa\lambda} & = \delta_\lambda L^\alpha_{\beta\kappa} - \delta_\kappa L^\alpha_{\beta\lambda} + L^\gamma_{\beta\kappa}L^\alpha_{\gamma\lambda} - L^\gamma_{\beta\lambda}L^\alpha_{\gamma\kappa} + C^\alpha_{\beta\gamma}\Omega^\gamma_{\kappa\lambda} \\
		P^\mu_{\nu\kappa\gamma} & = \pdot{\gamma} L^\mu_{\nu\kappa} - \mathcal D_\kappa C^\mu_{\nu\gamma} + C^\mu_{\nu\beta} \mathcal T^\beta_{\kappa\gamma} \\
		P^\alpha_{\beta\kappa\gamma} & = \pdot{\gamma} L^\alpha_{\beta\kappa} - \mathcal D_\kappa C^\alpha_{\beta\gamma} + C^\alpha_{\beta\delta} \mathcal T^\delta_{\kappa\gamma} \\
		S^\mu_{\nu\gamma\delta} & = \pdot{\delta} C^\mu_{\nu\gamma} - \pdot{\gamma}C^\mu_{\nu\delta} + C^\kappa_{\nu\gamma}C^\mu_{\kappa\delta} - C^\kappa_{\nu\delta}C^\mu_{\kappa\gamma} \\
		S^\alpha_{\beta\gamma\delta} & = \pdot{\delta} C^\alpha_{\beta\gamma} - \pdot{\gamma}C^\alpha_{\beta\delta} + C^\epsilon_{\beta\gamma}C^\alpha_{\epsilon\delta} - C^\epsilon_{\beta\delta}C^\alpha_{\epsilon\gamma}
	\end{align}
	where $\mathcal D_\kappa$ is the covariant derivative with respect to the connection defined in \eqref{metric d-connection 1} - \eqref{metric d-connection 4}. An explicit calculation for the SFR metric yields
	\begin{equation}
		R^\alpha_{\beta\kappa\lambda} = P^\mu_{\nu\kappa\gamma} = P^\alpha_{\beta\kappa\gamma} = S^\mu_{\nu\gamma\delta} = 0
	\end{equation}
	Consequently, we cannot construct any non-vanishing invariant of the metrics from these components so they are not useful for finding singular points. Additionally, we get $g^{\mu\nu}R_{\mu\nu} = 0$ so this scalar curvature gives no information about singular points either.
	
	Next, we calculate the scalar curvature $S = v^{\alpha\beta}S_{\alpha\beta}$ to the lowest non-vanishing order and we find:
	\begin{equation}\label{ScurvatureSFR}
		S = \frac{5 \tilde A_0^2 r \left\{ (y^r)^2 + r (r - R_S) \left[ (y^\theta)^2 + \sin^2\theta (y^\phi)^2 \right] \right\}}{2 \tilde a^4 \left(r - R_S\right)}
	\end{equation}
	with $\tilde{a} = \sqrt{-g_{\alpha\beta}y^{\alpha}y^{\beta}}$ and we have set $y^4 \equiv y^t, y^5 \equiv y^r, y^6 \equiv y^\theta, y^7 \equiv y^\phi$. From \eqref{ScurvatureSFR}, we see that the anisotropic scalar curvature $ S $ has a geometrical meaning because of its dependence on the coordinates.
	
	A straightforward calculation results in the following cases:
	\begin{enumerate}
		\item $\tilde a \neq 0$, $r = R_S$ and $y^r \neq 0$: In this case, we get $S=0$ 
		\item $\tilde a \neq 0$, $r = R_S$ and $y^r = 0$: In this case, the fiber scalar curvature takes the value 
		\begin{equation}
			S = \frac{5 \tilde A_0^2}{2 R_S^2\left[ (y^\theta)^2 + \sin^2\theta (y^\phi)^2 \right] }
		\end{equation}
		\item $\tilde a = 0$: In this case, the fiber scalar curvature diverges: $S \rightarrow \infty$ 
	\end{enumerate}
	The third case is the most interesting one, where it can be seen that $\tilde a = 0$ represents a set of singular points for the metric $v_{\alpha\beta}$. In the next paragraphs, we will identify $y^\alpha$ with the 4-velocity of a free particle, in which case the condition $\tilde a = 0$ will denote a null path with respect to the metric $g_{\mu\nu}(x)$. Taking this argument into account, we reach the conclusion that such paths can not describe physical trajectories.
	
	Finally, we calculate the nontrivial Kretschmann-like invariants of the metrics $g_{\mu\nu}$ and $v_{\alpha\beta}$ to the lowest non-vanishing order:
	\begin{align}
		K_H & \equiv R_{\kappa\lambda\mu\nu}R^{\kappa\lambda\mu\nu} = \frac{12 R_S^2}{r^6} \label{kretschmann h} \\
		K_V & \equiv S_{\alpha\beta\gamma\delta}S^{\alpha\beta\gamma\delta} = \left( \frac{3 S}{5} \right)^2 \label{kretschmann v}
	\end{align}
	The invariant in Eq.\eqref{kretschmann h} coincides with the Kretschmann invariant of the classic Schwarzschild solution \cite{Carroll} and it reveals a singularity of the metric $g_{\mu\nu}$ at the point $r=0$. The second Kretschmann-like invariant contains the same information as the scalar curvature $S$, as we can see from Eqs.\eqref{ScurvatureSFR} and \eqref{kretschmann v}, so the same conclusions apply for it.
	
	We notice from \eqref{kretschmann h} and \eqref{kretschmann v} that the total Kretschmann invariant $K = K_H + K_V$ is equal to the classic Schwarzschild one plus a small correction which comes from the additional geometrical inner structure of the SFR gravitational model. Specifically, the scalar curvature of the vertical space (the space of $y-$variables) is related to a non-trivial vertical-space Kretschmann invariant, as one can see from \eqref{kretschmann v}, so it induces a deviation from classical general relativity.
	
	\section{Paths}\label{sec: Paths}
	In this section, we study the paths of a particle in the SFR model. We consider the Lagrangian of the form \cite{Triantafyllopoulos:2020ogl}:
	\begin{equation}\label{full lagrangian}
		L(x,\dot x,y) = \left(-a g_{\mu\nu}\dot x^\mu \dot x^\nu - b \tilde \delta^\alpha_\mu v_{\alpha\beta}\dot x^\mu y^\beta - c v_{\alpha\beta}y^\alpha y^\beta \right)^{1/2}
	\end{equation}
	with $a, b, c$ constants.
	Variation of the action with respect to $y^\alpha$ gives the relation:
	\begin{equation}\label{y=xdot}
		y^\alpha = \dot x^\alpha
	\end{equation}
	Furthermore, if we variate the action with respect to $x^\mu$ and substitute \eqref{y=xdot}, we get the path equations:
	\begin{align}\label{timelike_path}
		& \ddot x^\mu + \gamma^\mu_{\kappa\lambda} \dot x^\kappa \dot x^\lambda \nonumber \\
		& = -\frac{z}{1+z}  \Big\{ \tilde a g^{\mu\nu} \left(\partial_\nu A_\kappa - \partial_\kappa A_\nu \right) \dot x^\kappa \nonumber\\
		& + \frac{1}{\tilde a} \Big[A^\nu \left(\partial_\kappa g_{\nu\lambda} - \frac{1}{2}\partial_\nu g_{\kappa\lambda} \right) + \partial_\kappa A_\lambda \Big] \dot x^\mu \dot x^\kappa \dot x^\lambda \nonumber\\
		& + \frac{1}{\tilde a} \Big( \frac{1}{4}g^{\mu\nu}\! A_\kappa \partial_\nu g_{\sigma\lambda} + g^{\mu\nu}\! A_\kappa \partial_\lambda g_{\sigma\nu} + A^\mu \partial_\kappa g_{\lambda\sigma} \Big) \dot x^\sigma \dot x^\kappa \dot x^\lambda \nonumber \\
		& + \frac{1}{2\tilde{a}^3} A_\lambda \partial_\kappa g_{\sigma\tau} \dot x^\sigma \dot x^\tau \dot x^\mu \dot x^\kappa \dot x^\lambda \Big\}
	\end{align}
	where $z = -b^2/4ac$ is a constant and a dot denotes differentiation with respect to the generalized proper time $\tau$, with the definition
	\begin{equation}\label{pathmetric}
		d\tau = \left[-a g_{\mu\nu} dx^\mu dx^\nu - ( b + c) v_{\alpha\beta}dx^\alpha dx^\beta \right]^{1/2}
	\end{equation}
	which is derived from the Lagrangian \eqref{full lagrangian} if we substitute $y^\alpha = dx^\alpha$. The form \eqref{timelike_path} generalizes the geodesics equations of general relativity in the SFR model.
	
	In order to solve the equations \eqref{timelike_path} we use the NDSolve command of Mathematica to obtain a numerical timelike solution. By assuming different initial values we get two different solutions which are described by a closed path and an open path respectively and we compare our results with the geodesics of GR for the same initial values. In our approach, we consider the energy $E= \left(1-\frac{R_s}{r}\right)\der{t}{\tau}$ and the angular momentum $L=r^{2}\der{\phi}{\tau}$ .
	\begin{figure}[ht]
		\centering
		\includegraphics[scale=0.4]{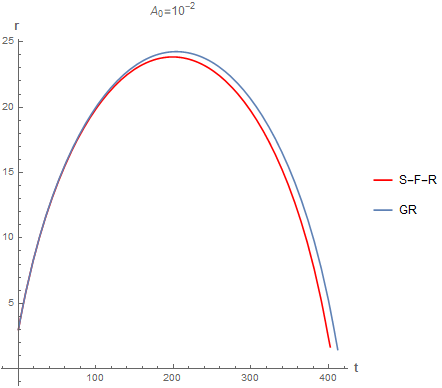}
		\caption{This is an $r,t$ graph of the timelike paths that we find using our theoretical SFR (red curve) 
			model in comparison to the geodesics of GR (blue line) for $E=0.98$, $L=1$, $r_0=3$ and $(a,b,c)=(1,1,1)$.}
		\label{closed1 r-t}
	\end{figure}
	\begin{figure}[ht]
		\centering
		\includegraphics[scale=0.4]{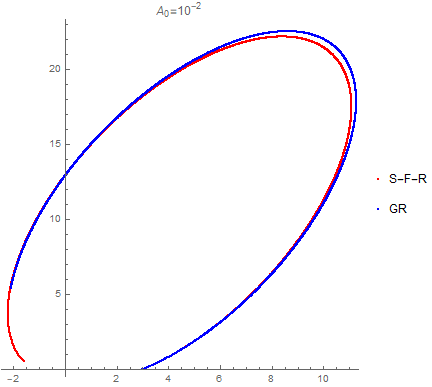}
		\caption{This is a polar graph of the timelike paths that we find using our theoretical SFR (red curve) 
			model in comparison to the geodesics of GR (blue line) for $E=0.98$, $L=1$, $r_0=3$ and $(a,b,c)=(1,1,1)$.}
		\label{closed1 polar}
	\end{figure}
	
	We notice from the two graphs (Fig.\ref{closed1 r-t} and Fig.\ref{closed1 polar}) that the paths in the SFR model and GR are very similar. However, from the r-t graph (Fig.\ref{closed1 r-t}) we can see that the maximum radial distance in SFR is lower and the required time to reach the Schwarzschild radius is also less compared to GR. From the second graph (Fig.\ref{closed1 polar}) we can see that the two ellipses are similar but the red ellipse (SFR model) is smaller and it reaches the event horizon faster than the blue ellipse (GR). We remark that in the path equations \eqref{timelike_path} the right hand side is non-zero and this term acts as a small extra force that influences the paths in the gravitational field. This correction increases or decreases the effects of gravity depending on the sign of the term.
	\begin{figure}[ht]
		\centering
		\includegraphics[scale=0.15]{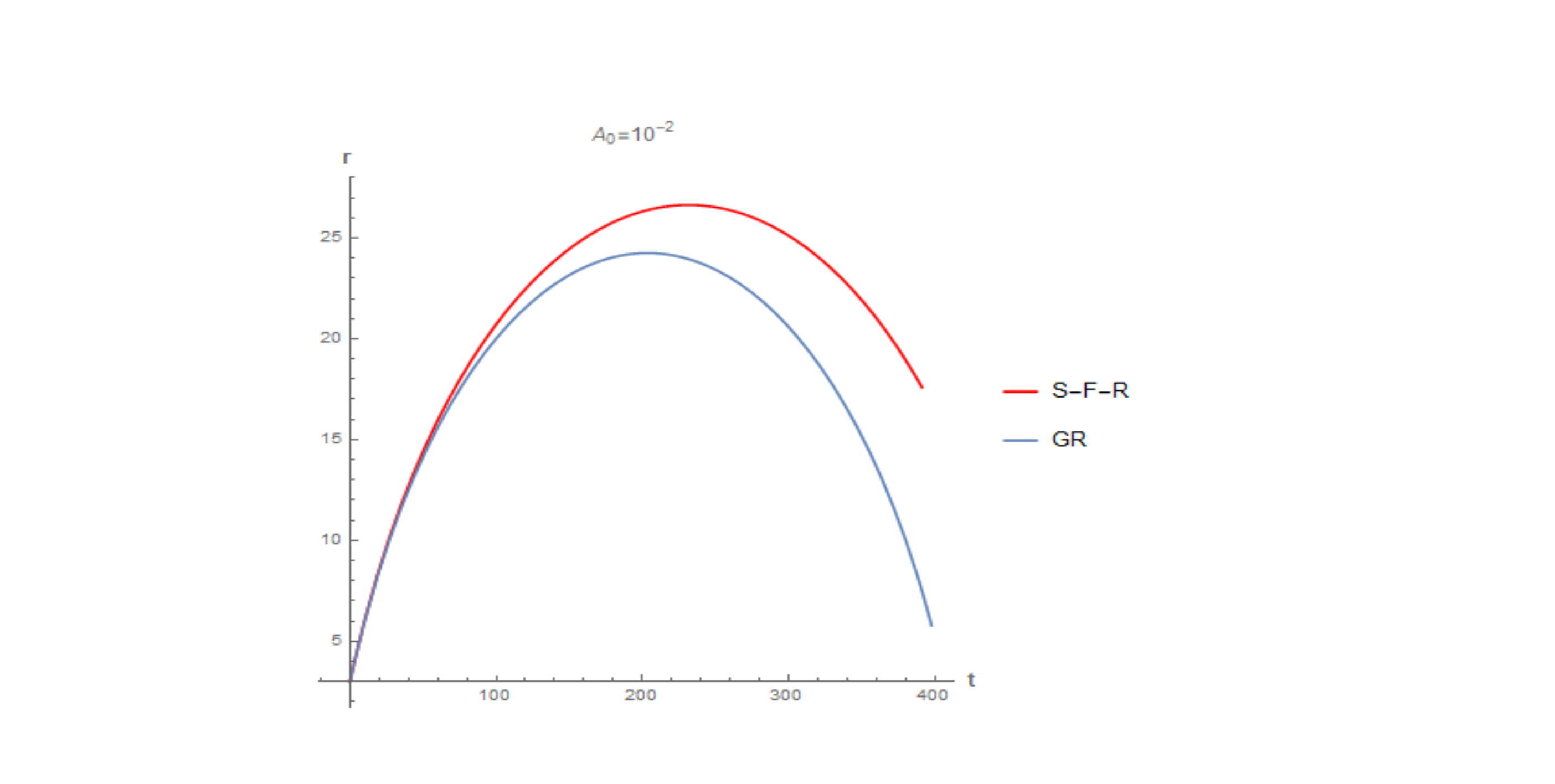}
		\caption{This is an $r,t$ graph of the timelike paths that we find using our theoretical SFR (red curve) 
			model in comparison to the geodesics of GR (blue line) for $E=0.98$, $L=1$, $r_0=3$ and $(a,b,c)=(1,10,10)$.}
		\label{closed2 r-t}
	\end{figure}
	\begin{figure}[ht]
		\centering
		\includegraphics[scale=0.4]{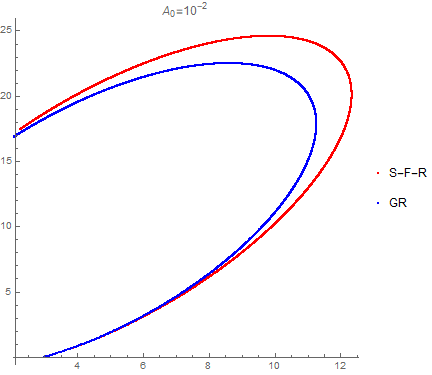}
		\caption{This is a polar graph of the timelike paths that we find using our theoretical SFR (red curve) 
			model in comparison to the geodesics of GR (blue line) for $E=0.98$, $L=1$, $r_0=3$ and $(a,b,c)=(1,10,10)$.}
		\label{closed2 polar}
	\end{figure}
	
	In the two figures (Fig.\ref{closed2 r-t} and Fig.\ref{closed2 polar}) we have taken $a=1$, $b=10$ and $c=10$ in \eqref{full lagrangian}. In this case, we can see that  the red line (SFR model) takes higher values than the blue line (GR) and it requires more time to reach the Schwarzschild radius. In our case, the parameters $(\tilde{A}_{0}, a, b, c)$ control the deviation of the SFR model from General Relativity. In particular, the values of $(a,b,c)$ can give higher or lower results compared to GR.
	
	\begin{figure}[ht]
		\centering
		\includegraphics[scale=0.4]{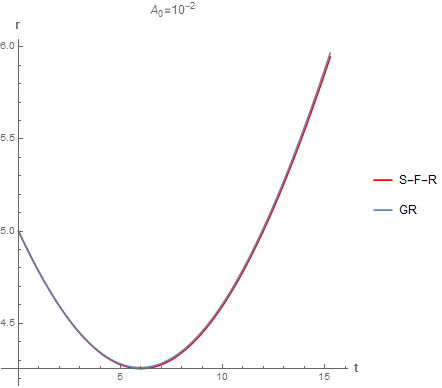}
		\caption{This is an r-t graph of the timelike paths in the SFR model in comparison to the geodesics in GR for $E=1.2$, $L=4$, $r_0=5$ and $(a,b,c)=(1,10,10)$.}
		\label{open1 r-t}
	\end{figure}
	
\begin{figure}[ht]
		\centering
		\includegraphics[scale=0.4]{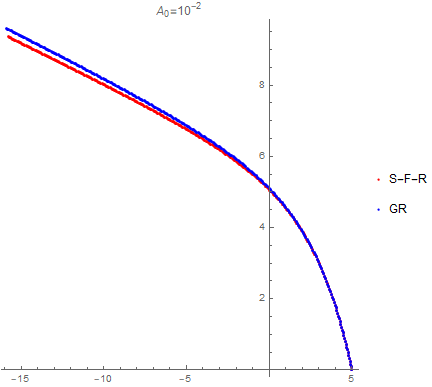}
		\caption{This is a polar graph of the timelike paths in the SFR model in comparison to the geodesics in GR for $E=1.2$, $L=4$, $r_0=5$ and $(a,b,c)=(1,10,10)$.}
		\label{open1 polar}
	\end{figure}

	The last two graphs (Fig.\ref{open1 r-t} and Fig.\ref{open1 polar}) represent open paths with $a=1$, $b=10$ and $c=10$. In this case, we can see that the SFR model deviates from GR when we start to move away from the event horizon and the two paths (red and blue) separate.
	For a small interval, the paths of the SFR model approximate the geodesics of GR. As the radial distance increases, the paths of our model deviate from GR.\\  
	\section{Energy}\label{sec: Energy}
	In this section, we give the form of the energy and momentum of a particle in an SFR spacetime.
	
	We assume a four-velocity vector $u^{\alpha}=(u^{t},u^{r},u^{\theta},u^{\phi})$, with
	\begin{equation}
		u^\alpha \equiv \der{x^\alpha}{\tau}
	\end{equation}
	and we require that its norm equals $-1$, so we have \cite{Relancio:2020mpa}:
	\begin{equation}
		||u|| = u^{\alpha}u_{\alpha} =
		u^{\alpha}u^{\beta}v_{\alpha\beta} = -1
	\end{equation}
	By use of \eqref{vab} we find\footnote{The condition \eqref{u norm} along with relation \eqref{pathmetric} give $a=b+c=1$ in this case.}:
	\begin{equation}\label{u norm}
		g_{\alpha\beta}u^{\alpha}u^{\alpha} + w_{\alpha\beta}u^{\alpha}u^{\beta} = -1    
	\end{equation}
	and by \eqref{wab} we get:
	\begin{align}
		&g_{\alpha\beta}u^{\alpha}u^{\beta} + \frac{1}{\tilde a}\Big[(A_{\beta}u^{\beta}) (g_{\alpha\gamma}u^{\alpha}u^{\gamma}) \nonumber\\ &+(A_{\gamma}u^{\gamma})(g_{\alpha\beta}u^{\alpha}u^{\beta}) +(A_{\alpha}u^{\alpha}(g_{\beta\gamma}u^{\beta}u^{\gamma})\Big]\nonumber\\ 
		&+ \frac{1}{\tilde a^{3}}(A_{\gamma}u^{\gamma})(g_{\alpha\epsilon}u^{\alpha}u^{\epsilon})(g_{\beta\delta}u^{\beta}u^{\delta}) = -1   
	\end{align}
	where we have set $y^\alpha = u^\alpha$ and $\tilde{a} = \sqrt{-g_{\alpha\beta}u^{\alpha}u^{\beta}}$.
	
	After some calculations, we have:
	\begin{equation}\label{eqnorm}
		\tilde a^{2} + 2A_{\gamma}u^{\gamma}\tilde{a} -1 = 0
	\end{equation}
	By solving \eqref{eqnorm} we can find $\tilde a$ 
	\begin{equation}
		\tilde a = -\tilde A_{0}f^{1/2}u^{t} + \sqrt{1 + \tilde A_{0}^{2}f(u^{t})^{2}}    
	\end{equation}
	where we have used \eqref{Asolution} for $A_{\gamma}$ with $f = 1-\frac{R_S}{r}$.\\[7pt]
	If we use a Taylor expansion for the second term and omit higher order terms $O(\tilde{A_{0}}^2)$ we get:
	\begin{equation}\label{eqnorma}
		\tilde a = 1 - \tilde A_{0}f^{1/2}u^{t}    
	\end{equation}
	Equation \eqref{eqnorma} is the condition so that the norm of the four-velocity equals $-1$. 
	
	If we assume that the particle is at rest, the four-velocity becomes $u^\alpha = (u^{t},0,0,0)$ and if we substitute this in \eqref{eqnorma} we find:
	\begin{equation}\label{utSFR}
		u^{t}_{SFR} = (1-\tilde A_{0})f^{-1/2}    
	\end{equation}
	We see from \eqref{utSFR} that if $\tilde A_0 \rightarrow 0$ we find the result from GR :
	\begin{equation}\label{utGR}
		u^{t}_{GR} = f^{-1/2}    
	\end{equation}
	Consequently, by using \eqref{utSFR} and \eqref{utGR} we can write:
	\begin{equation}\label{utcomp}
		u^{t}_{SFR} = (1-\tilde A_{0})u^{t}_{GR}    
	\end{equation}
	From rel.\eqref{utcomp} we see that if $\tilde A_{0}$ has a positive value then $u^{t}_{SFR}<u^{t}_{GR}$ and if
	$\tilde A_{0}$ has a negative value then $u^{t}_{SFR}>u^{t}_{GR}$.\\[7pt]
	We can find the momentum and energy of the particle:
	\begin{equation}\label{momentumsfr}
		p^{\alpha}=mu^{\alpha}=(mu^{t},0,0,0)
	\end{equation}
	where m is the mass of the particle.
	\\
	From rel.\eqref{utSFR} we get for $p^{t}_{SFR}$ and $E_{SFR}$:
	\begin{equation}\label{energysfr}
		E_{SFR}=p^{t}_{SFR}=m(1-\tilde A_{0})f^{-1/2}    
	\end{equation}
	\section{Gravitational Redshift}\label{sec: Redshift}
	If we take $r,\theta,\phi=const$ in the definition of proper time \eqref{pathmetric}, we get:
	\begin{equation}
		d\tau = \left[ -a g_{00}dt^2 - ( b + c)v_{00}dt^2 \right]^{1/2}    
	\end{equation}
	By using eq.\eqref{vab}, we get:
	\begin{equation}\label{dtau}
		d\tau = [-g_{00} - \kappa w_{00}]^{1/2}dt'    
	\end{equation}
	where we have set $dt'=\sqrt{a+b+c}dt$ and $\kappa=\frac{b+c}{a+b+c}$.\\[7pt]
	From the definition of the metric perturbation $w_{\alpha\beta}$ in \eqref{wab} for $\alpha =0$ and $\beta=0$ we get:
	\begin{align}
		w_{00} = & \, \frac{1}{\tilde a} \left(A_{0}g_{00}\dot{x}^{0}+A_{0}g_{00}\dot{x}^{0}+A_{0}g_{00}\dot{x}^{0} \right) \nonumber\\
		& +\frac{1}{\tilde a^{3}}A_{0}g_{00}g_{00}\dot{x}^{0}\dot{x}^{0}\dot{x}^{0} \nonumber\\
		\Rightarrow w_{00} = & \, -2\tilde A_{0}f
	\end{align}
	where $\tilde a = \sqrt{-g_{\alpha\beta}\dot x^{\alpha}\dot x^{\beta}}=\dot t\sqrt{-g_{00}} = \dot t\sqrt{f}$ because we have taken $r,\theta,\phi=$const.\\[7pt]
	
	If we return to \eqref{dtau} we find:
	\begin{align}
		d\tau &= [-g_{00} - \kappa w_{00}]^{1/2}dt'\nonumber\\[7pt]
		\Rightarrow	d\tau &=[f-\kappa(-2\tilde A_{0}f)]^{1/2}dt'\nonumber\\[7pt]
		\Rightarrow	d\tau &= (1+\epsilon)^{1/2}\sqrt{-g_{00}}\,dt'
	\end{align}
	where we have set $\epsilon=2\kappa\tilde A_{0}$ with $\epsilon \ll 1$, $g_{00}=-f$ and $f=1-\frac{R_s}{r}$. We note that in GR the calculation for the redshift leads to : $d\tau_{GR}=\sqrt{-g_{00}}\,dt$.
	
	Now, if we consider two clocks at two different points of spacetime $r_1$ and $r_2$, we will have:
	\begin{equation}
		d\tau_1 = (1+\epsilon)^{1/2}\sqrt{-g_{00}(1)}\,dt'
	\end{equation}
	and
	\begin{equation}
		d\tau_2 = (1+\epsilon)^{1/2}\sqrt{-g_{00}(2)}\,dt'   
	\end{equation}
	and thus for the frequencies $\nu_1$ and $\nu_2$ we find:
	\begin{align}
		\nu_2 & = \nu_1\left(\frac{g_{00}(1)}{g_{00}(2)}\right)^{1/2} = \nu_1\left(\frac{1-\frac{R_s}{r_1}}{1-\frac{R_s}{r_2}}\right)^{1/2}\nonumber   \\[7pt]
		\Rightarrow\frac{\nu_2}{\nu_1} & \approx 1 - GM \left(\frac{1}{r_1}-\frac{1}{r_2}\right)   \label{frequencies quotient}
	\end{align}
	where we have used the Taylor expansion $(1+x)^{1/2}\approx 1+\frac{1}{2}x$.\\[7pt]
	From \eqref{frequencies quotient} we find :
	\begin{equation}\label{sfrredshift}
		\left(\frac{\Delta \nu}{\nu_1}\right)_{SFR} = \Delta U   
	\end{equation}
	where $\Delta\nu=\nu_{2}-\nu_{1}$ with $\nu_2,\nu_1$ the emitter and receiver frequencies and $\Delta U=GM(\frac{1}{r_2}-\frac{1}{r_1})$ is the change of potential.
	
	We recall that in general relativity (GR) the gravitational redshift is given by:
	\begin{equation}\label{grredshift}
		\left(\frac{\Delta\nu}{\nu_{1}}\right)_{GR}=\Delta U
	\end{equation}
	We remark that, in the scenario under consideration, the gravitational redshift predicted by the SFR gravitational model is the same as the one predicted in the classic Schwarzschild spacetime of GR.

	\section{Photonsphere}\label{sec: Photonsphere}
	In order to calculate the radius of the photonsphere we will use eq.\eqref{pathmetric}:
	\begin{equation}
		d\tau = \left[-a g_{\mu\nu} dx^\mu dx^\nu - ( b + c) v_{\alpha\beta}dx^\alpha dx^\beta \right]^{1/2}
	\end{equation}
	From \eqref{vab} we get:
	\begin{equation}\label{pathmetric2}
		d\tau'=\left(-g_{\mu\nu}dx^{\mu}dx^{\nu}-\kappa w_{\alpha\beta}dx^{\alpha}dx^{\beta}\right)^{1/2}
	\end{equation}
	where $\kappa=\frac{b+c}{a+b+c}$ and $d\tau'=\frac{d\tau}{\sqrt{a+b+c}}$.\\[7pt]
	To calculate the radius of the photonsphere, we take $r=$ const. , $\theta=\frac{\pi}{2}$ and $d\tau'=0$ because we want to find the photon orbits.\\
	Under these conditions, rel.\eqref{pathmetric2} yields:
	\begin{equation}
		(g_{00}+\kappa w_{00})dt^{2}+(g_{33}+\kappa w_{33})d\phi^{2}=0    
	\end{equation}
	From the above relation, we find:
	\begin{equation}\label{photoneq1a}
		\left(\frac{d\phi}{dt}\right)^{2}=-\frac{g_{00}+\kappa w_{00}}{g_{33}+\kappa w_{33}}   
	\end{equation}
	To calculate $w_{00}$ and $w_{33}$, we use \eqref{wab}.
	\begin{align}
		w_{\alpha\beta} = & \, \frac{1}{\tilde{a}}(A_{\beta}g_{\alpha\gamma}y^\gamma + A_{\gamma}g_{\alpha\beta}y^\gamma + A_{\alpha}g_{\beta\gamma}y^\gamma) \nonumber\\
		& + \frac{1}{\tilde{a}^3}A_{\gamma}g_{\alpha\epsilon}g_{\beta\delta}y^\gamma y^\delta y^\epsilon
	\end{align}
	where $\tilde{a}=\sqrt{-g_{\alpha\beta}y^{\alpha}y^{\beta}}$ and for $A_{\gamma}$ we use \eqref{Asolution}.\\[7pt]
	We calculate $\tilde a$ :
	\begin{align}\label{alphaphoton}
		\tilde{a}=\sqrt{-g_{\alpha\beta}y^{\alpha}y^{\beta}}&=\sqrt{-g_{\alpha\beta}\dot x^{\alpha}\dot x^{\beta}}=\sqrt{f\dot t^{2}-r^{2}\dot\phi^{2}}\Rightarrow\nonumber\\[7pt]
		\tilde{a}=&\dot t\sqrt{f-r^{2}\phi'^{2}}=\dot t\tilde p
	\end{align}
	where we used the Leibniz chain rule $\dot\phi=\frac{d\phi}{dt}\frac{dt}{d\tau}=\phi'(t)\dot t$ and we set $\tilde p=\sqrt{f-r^2\phi'^{2}}$.\\[7pt]
	After some calculations, we find:
	\begin{align}
		w_{00}&=\frac{\tilde A_{0}}{\tilde p^{3}}f^{3/2}\big(-3\tilde a^{2} + f\big)\\[7pt]   
		w_{33}&=\frac{\tilde A_{0}}{\tilde p^{3}}f^{3/2}r^{2}
	\end{align}
	If we return to \eqref{photoneq1a} and we use $w_{00}$ and $w_{33}$, we get:
	\begin{equation}\label{dsequation}
		r^2\phi'^2+\kappa\frac{\tilde A_{0}}{\tilde p^{3}}f^{3/2}r^2\phi'^2 = f-\kappa\frac{\tilde A_{0}}{\tilde p^{3}}f^{3/2}(f-3\tilde a^2)    
	\end{equation}
	From \eqref{dsequation} we find:
	\begin{equation}\label{photoneq1b}
		\tilde p^5 + 4\kappa\tilde A_{0}f^{3/2}\tilde p^2 - 2\kappa\tilde A_{0}f^{5/2}=0    
	\end{equation}
	In order to determine the radius of the photonsphere, we need two equations. The first one is \eqref{photoneq1b} and we find the second from the path equations. We get the radial path equation by substituting $\mu=1$ in \eqref{timelike_path} and if we use our assumptions $r=$const. and $\theta=\frac{\pi}{2}$ we find:
	\begin{align}
		&\frac{f(1-f)}{2r} \dot t^2 - rf\dot\phi^2=\nonumber \\
		&-\lambda\tilde A_{0}\left[\left(\frac{1}{2}\tilde{a}f^{1/2}\dot t - \frac{1}{4\tilde{a}}f^{3/2}\dot t^3\right)\frac{1-f}{r}+ \frac{1}{2\tilde{a}}f^{3/2}r\dot t\dot\phi^2\right]
	\end{align}
	where $\lambda=\frac{z}{1+z}$.\\[7pt]
	Then, by using \eqref{alphaphoton} and after some calculations we find:
	\begin{equation}\label{photoneq2}
		4f^{1/2}\tilde p^{3} + 2\lambda\tilde A_{0}(1-2f)\tilde p^2 + 2f^{1/2}(1-3f)\tilde p -\lambda\tilde A_{0}f(1-3f)=0    
	\end{equation}\\[7pt]
	Therefore, the equations we need to solve are \eqref{photoneq1b} and \eqref{photoneq2}. If we take \eqref{photoneq1b} and set $\mu= f^{-1/2}\tilde p$ we get:
	\begin{equation}\label{mueq}
		\mu^5 + 4\kappa\tilde A_{0}\mu^2 - 2\kappa\tilde A_{0}=0    
	\end{equation}
	By giving values to the parameters $\kappa$ and $\tilde A_{0}$ we can solve \eqref{mueq} numerically and determine the value of $\mu$. Then, from the definition of $\mu$ we can find a relation between $f$ and $\tilde p$ which can be substituted in \eqref{photoneq2} to find the term $f$ and from this the radius of the photonsphere. The results for different values of the parameters are shown on the table that follows:
	\begin{center}
		\begin{tabular}{|c|c|c|c|}
			\hline
			($a$ , $b$ , $c$) & $\tilde A_{0}$ & $\mu$ &$r/R_s$\\[1ex]
			\hline\hline
			$(1,1,1)$ & $10^{-3}$ & $0,25854$ & $1,53577$ \\[2ex]
			\hline
			$(1,1,1)$ & $10^{-4}$ & $0,16599$ & $1,51416$ \\[2ex]
			\hline
			$(1,1,1)$ & $10^{-6}$ & $0,06671$ & $1,50224$ \\[2ex]
			\hline
			$(1000,1,1)$ & $10^{-4}$ & $0,05245$ & $1,50138$ \\[2ex]
			\hline
			$(1,1000,1)$ & $10^{-4}$ & $0,17961$ & $1,51668$ \\[2ex]
			\hline
			$(1,1,1000)$ & $10^{-4}$ & $0,17961$ & $1,51667$ \\[2ex]
			\hline
			$(1,1000,1000)$ & $10^{-4}$ & $0,17963$ & $1,51668$ \\[2ex]
			\hline
		\end{tabular}    
	\end{center}
	where $a,b,c$ are the starting parameters in the Lagrangian in \eqref{full lagrangian} and through them we calculate the term $\kappa = \frac{b+c}{a+b+c}$.
	\section{Concluding remarks}\label{sec: Conclusion}
	In this article, we investigate further properties and applications of our previous work of the SFR model which generalizes the classical Schwarzschild spacetime by introducing a timelike covector $A_{\gamma}$ in the metric structure \cite{Triantafyllopoulos:2020vkx}. This covector is specified by the solution of the generalized Einstein equations of the SFR model. It provides the local anisotropy and may cause Lorentz violating effects. 
	
	In addition, we derive the form of S-anisotropic curvature which takes a geometrical meaning because of its dependence on coordinates.
	
	The generalized Kretschmann-like curvature invariant plays a crucial role in our approach since the horizontal $K_{H}$, rel.\eqref{kretschmann h}, coincides with the Kretschmann invariant of the classical Schwarzschild solution which gives a singularity at the point $r=0$. The second Kretschmann curvature invariant $K_{V}$, rel.\eqref{kretschmann v}, provides information for singularities with more degrees of freedom as we show and it is characterized by the scalar curvature S, rel.\eqref{ScurvatureSFR}.
	
	In the framework of applications of SFR model we extend our study of timelike geodesic paths and we compare them with corresponding paths of GR. We notice that the extra terms in rel.\eqref{timelike_path} act as an extra force that influences the gravitational field and give a small deviation from the paths of GR.
	
	In the last sections, we find the form of momentum and the energy in our approach, rel.\eqref{momentumsfr} and \eqref{energysfr}.
	
	By considering the Lagrangian function (rel.\eqref{full lagrangian}) we calculate the gravitational redshift and the photonsphere for our case. While in the redshift calculation we find no deviation from general relativity, in the study of the photonsphere we find infinitesimal deviations from GR which may be ought to the small anisotropic perturbations coming from Lorentz violation effects.
	
	\appendix
	\section{Variational principle on a Hilbert-like action}\label{sec: Variation}
	In this section, we present the basic steps of the variation of the action \eqref{Hilbert like action}:
		\begin{equation}
			K = \int_{\mathcal N} d^8\mathcal U \sqrt{|\Gd|}\, \R + 2\kappa \int_{\mathcal N} d^8\mathcal U \sqrt{|\Gd|}\,\mathcal L_M
		\end{equation}
		with respect to $g_{\mu\nu}$, $v_{\alpha\beta}$ and $N^\alpha_\kappa$ in order to acquire the generalized field equations \eqref{feq1}-\eqref{feq3}, see \cite{Triantafyllopoulos:2020ogl} for the original derivation. Variating the total action, we get:
		\begin{align}
			\Delta K = &\, \int_{\mathcal N} d^8 \mathcal U ( R + S) \Delta \sqrt{|\Gd|} + \int_{\mathcal N} d^8 \mathcal U \sqrt{|\Gd|} (\Delta  R + \Delta S) \nonumber\\
			& +\,  2\kappa\int_{\mathcal N} d^8\mathcal U \,\Delta\!\left(\sqrt{|\Gd|}\,\mathcal L_M\right) \label{DK}
		\end{align}
		with
		\begin{align}
			\Delta\sqrt{|\Gd|} = & -\frac{1}{2}\sqrt{|\Gd|}\left( g_{\mu\nu}\Delta g^{\mu\nu} + v_{\alpha\beta}\Delta v^{\alpha\beta}\right) \label{DG}\lin
			\Delta R = & \, 2g^{\mu[\kappa}\pdot{\alpha}L^{\nu]}_{\mu\nu} \Delta N^\alpha_\kappa + \overline{R}_{\mu\nu}\Delta g^{\mu\nu} + \mathcal D_\kappa Z^\kappa \label{DR} \lin
			\Delta S = &\, S_{\alpha\beta} \Delta v^{\alpha\beta} + \mathcal D_\gamma B^\gamma \label{DS}
		\end{align}
		where $\overline R_{\mu\nu} = R_{(\mu\nu)} + \Omega^\alpha_{\kappa(\mu} C^\kappa_{\nu)\alpha}$ and
		\begin{align}
			Z^\kappa = &\, g^{\mu\nu}\Delta L^\kappa_{\mu\nu} - g^{\mu\kappa} \Delta L^\nu_{\mu\nu} \nonumber\lin
			= &\, -\D_\nu\Delta g^{\nu\kappa} + g^{\kappa\lambda}g_{\mu\nu}\D_\lambda \Delta g^{\mu\nu} \nonumber\lin
			& + \, 2\left( g^{\kappa\mu}C^\lambda_{\lambda\alpha} - g^{\kappa\lambda}C^\mu_{\lambda\alpha} \right) \Delta N^\alpha_\mu \label{Zdef}\lin
			B^\gamma = &\, v^{\alpha\beta}\Delta C^\gamma_{\alpha\beta} - v^{\alpha\gamma}\Delta C^\beta_{\alpha\beta} \nonumber\lin
			= &\, -\D_\alpha \Delta v^{\alpha\gamma} + v^{\gamma\delta}v_{\alpha\beta} \D_\delta\Delta v^{\alpha\beta} \label{Bdef}
		\end{align}
		
		Stokes theorem on the Lorentz tangent bundle reads:
		\begin{align}
			\int_{\mathcal N} d^8 \mathcal{U}\sqrt{|\Gd|}\,\mathcal D_\mu H^\mu = & \int_{\mathcal N} d^8 \mathcal{U}\sqrt{|\Gd|}\,\mathcal T^\alpha_{\mu\alpha}H^\mu \nonumber\\
			& + \, \int_{\partial {\mathcal N}} n_\mu H^\mu \mathcal{\tilde E} \label{stokesh}\\
			\int_{\mathcal N} d^8 \mathcal{U}\sqrt{|\Gd|}\,\mathcal D_\alpha W^\alpha = &  -\int_{\mathcal N} d^8 \mathcal{U}\sqrt{|\Gd|}\, C^\mu_{\mu\alpha}W^\alpha \nonumber\\ 
			& + \, \int_{\partial {\mathcal N}}n_\alpha W^\alpha \mathcal{\tilde E} \label{stokesv}
		\end{align}
		where $H = H^\mu \delta_\mu$ and $W = W^\alpha \pdot{\alpha}$ are vector fields on $TM$, $\tilde{\mathcal E}$ is the Levi-Civita tensor on the boundary $\partial {\mathcal N}$, $(n_\mu,n_\alpha)$ is the normal vector on the boundary and $\mathcal T^\alpha_{\mu\beta} = \pdot{\beta}N^\alpha_\mu - L^\alpha_{\beta\mu}$. Using relation \eqref{stokesh} and eliminating boundary terms, we get
		\begin{align}
			&\int_{\mathcal N} d^8 \mathcal{U}\sqrt{|\Gd|}\,\mathcal D_\kappa Z^\kappa =   \int_{\mathcal N} d^8 \mathcal{U}\sqrt{|\Gd|}\,\mathcal T^\alpha_{\kappa\alpha}Z^\kappa \nonumber\\
			& = \int_{\mathcal N} d^8 \mathcal{U}\sqrt{|\Gd|}\,\D_\nu \left[ \mathcal T^\beta_{\kappa\beta} \left( -\Delta g^{\nu\kappa} + g^{\nu\kappa}g_{\mu\lambda}\Delta g^{\mu\lambda}\right)\right] \nonumber\\
			& \, -\int_{\mathcal N} d^8 \mathcal{U}\sqrt{|\Gd|}\,\left[ -\D_\nu\mathcal T^\beta_{\mu\beta} + g_{\mu\nu}\D^\lambda\mathcal T^\beta_{\lambda\beta}\right]\Delta g^{\mu\nu} \nonumber\\
			& \, + 2\int_{\mathcal N} d^8 \mathcal{U}\sqrt{|\Gd|}\,\mathcal T^\beta_{\kappa\beta}\left( g^{\kappa\mu}C^\lambda_{\lambda\alpha} - g^{\kappa\lambda}C^\mu_{\lambda\alpha}\right)\Delta N^\alpha_\mu
		\end{align}
		where we have used the Leibniz rule for the covariant derivative. Using \eqref{stokesh} again and eliminating the new boundary terms, we get
		\begin{align}
			&\int_{\mathcal N} d^8 \mathcal{U}\sqrt{|\Gd|}\,\mathcal D_\kappa Z^\kappa = \nonumber\\ 
			& \int_{\mathcal N} d^8 \mathcal{U}\sqrt{|\Gd|}\,\left(\delta^{(\lambda}_\nu\delta^{\kappa)}_\mu - g^{\kappa\lambda}g_{\mu\nu} \right)\left(\mathcal D_\kappa\mathcal T^\beta_{\lambda\beta} - \mathcal T^\gamma_{\kappa\gamma}\mathcal T^\beta_{\lambda\beta}\right) \Delta g^{\mu\nu} \nonumber\\
			& + \int_{\mathcal N} d^8 \mathcal{U}\sqrt{|\Gd|}\,4\mathcal{T}^\beta_{\kappa\beta}g^{\kappa[\mu}C^{\lambda]}_{\lambda\alpha} \Delta N^\alpha_\mu \label{DZ}
		\end{align}
		Similarly, using relation \eqref{stokesv} and eliminating the boundary terms, we get
		\begin{align}
			&\int_{\mathcal N} d^8\mathcal U \sqrt{|\Gd|}\, \D_\alpha B^\alpha =  -\int_{\mathcal N} d^8\mathcal U \sqrt{|\Gd|}\,C^\mu_{\mu\beta}B^\beta \nonumber\\
			& = -\int_{\mathcal N} d^8\mathcal U \sqrt{|\Gd|}\,\D_\alpha \left[ C^\mu_{\mu\beta}\Delta v^{\alpha\beta} - v^{\alpha\beta}v_{\gamma\delta}C^\mu_{\mu\beta} \Delta v^{\gamma\delta}\right] \nonumber\\ 
			& - \int_{\mathcal N} d^8\mathcal U \sqrt{|\Gd|}\,\left( \D_\alpha C^\mu_{\mu\beta} - v^{\gamma\delta}v_{\alpha\beta}\D_\gamma C^\mu_{\mu\delta}\right)\Delta v^{\alpha\beta}
		\end{align}
		where again we used the Leibniz rule. Applying \eqref{stokesv} again and eliminating the new boundary terms, we get
		\begin{align}
			& \int_{\mathcal N} d^8\mathcal U \sqrt{|\Gd|}\,\D_\alpha B^\alpha  \nonumber\\
			& = \int_{\mathcal N} d^8\mathcal U \sqrt{|\Gd|}\,\left(v^{\gamma\delta}v_{\alpha\beta} - \delta^{(\gamma}_\alpha\delta^{\delta)}_\beta \right)\left(\mathcal D_\gamma C^\mu_{\mu\delta} - C^\nu_{\nu\gamma}C^\mu_{\mu\delta} \right) \label{DB}
		\end{align}
		
		The matter part of the action is written as:
		\begin{align}
			& \int_{\mathcal N} d^8\mathcal U \,\Delta\!\left(\sqrt{|\Gd|}\,\mathcal L_M\right) \nonumber\\
			= & \, \int_{\mathcal N} d^8\mathcal U \sqrt{|\Gd|}\frac{1}{\sqrt{|\Gd|}}\frac{\Delta\!\left(\sqrt{|\Gd|}\,\mathcal L_M\right)}{\Delta g^{\mu\nu}}\Delta g^{\mu\nu} \nonumber\\
			& + \int_{\mathcal N} d^8\mathcal U \sqrt{|\Gd|}\frac{1}{\sqrt{|\Gd|}}\frac{\Delta\!\left(\sqrt{|\Gd|}\,\mathcal L_M\right)}{\Delta v^{\alpha\beta}}\Delta v^{\alpha\beta} \nonumber\\
			& + \int_{\mathcal N} d^8\mathcal U \sqrt{|\Gd|}\frac{1}{\sqrt{|\Gd|}}\frac{\Delta\!\left(\sqrt{|\Gd|}\,\mathcal L_M\right)}{\Delta N^\alpha_\kappa}\Delta N^\alpha_\kappa \label{EM variation}
		\end{align}
		
		Finally, combining equations \eqref{DK}-\eqref{Bdef}, \eqref{DZ}, \eqref{DB}, \eqref{EM variation} and setting $\Delta K = 0$, we get the equations \eqref{feq1}-\eqref{feq3} and the energy-momentum tensors \eqref{em1}-\eqref{em3}.

	\section*{Acknowledgements}
	The authors would like thank the unknown referee for his valuable comments and remarks.
	
	This research is co-financed by Greece and the European Union (European Social Fund-ESF) through the Operational Programme ``Human Resources Development, Education and Lifelong Learning'' in the context of the project ``Strengthening Human Resources Research Potential via Doctorate Research'' (MIS-5000432), implemented by the State Scholarships Foundation (IKY).

\end{document}